\RequirePackage{ifpdf}
\ifpdf 
\documentclass[pdftex]{sigma}
\else
\documentclass{sigma}
\fi

\DeclareMathOperator{\csch}{csch}
\DeclareMathOperator{\arccot}{arccot}
\DeclareMathOperator{\arccoth}{arccoth}

\begin{document}

\allowdisplaybreaks

\renewcommand{\PaperNumber}{046}

\FirstPageHeading

\numberwithin{equation}{section}

\newcommand{\alphab}{\boldsymbol{\alpha}}
\newcommand{\lambdab}{\boldsymbol{\lambda}}

\ShortArticleName{Position-Dependent Mass Schr\"odinger Equations}

\ArticleName{Point Canonical Transformation \\ versus Deformed Shape Invariance\\ for Position-Dependent Mass Schr\"odinger Equations}

\Author{Christiane QUESNE}

\AuthorNameForHeading{C.~Quesne}

\Address{Physique Nucl\'eaire Th\'eorique et Physique Math\'ematique,  Universit\'e Libre de
Bruxelles, \\ Campus de la Plaine CP229, Boulevard~du Triomphe, B-1050 Brussels,
Belgium}

\Email{\href{mailto:cquesne@ulb.ac.be}{cquesne@ulb.ac.be}}

\ArticleDates{Received February 09, 2009, in f\/inal form March 19,
2009; Published online April 15, 2009}

\Abstract{On using the known equivalence between the presence of a position-dependent mass (PDM) in the Schr\"odinger equation and a deformation of the canonical commutation relations, a method based on deformed shape invariance has recently been devised for genera\-ting pairs of potential and PDM for which the Schr\"odinger equation is exactly solvable. This approach has provided the bound-state energy spectrum, as well as the ground-state and the f\/irst few excited-state wavefunctions. The general wavefunctions have however remained unknown in explicit form because for their determination one would need the solutions of a~rather tricky dif\/ferential-dif\/ference equation. Here we show that solving this equation may be avoided by combining the deformed shape invariance technique with the point canonical transformation method in a novel way. It consists in employing our previous knowledge of the PDM problem energy spectrum to construct a constant-mass Schr\"odinger equation with similar characteristics and in deducing the PDM wavefunctions from the known constant-mass ones. Finally, the equivalence of the wavefunctions coming from both approaches is checked.}


\Classification{81Q05; 81Q60}

\section{Introduction}

There exists a wide variety of physical problems in which an ef\/fective mass depending on the position is of utmost relevance, such as ef\/fective interactions in nuclear physics \cite{ring}, carriers and impurities in crystals \cite{luttinger}, quantum dots \cite{harrison}, quantum liquids \cite{arias}, semiconductor heterostructures \cite{weisbuch}, and physics in neutron stars \cite{chamel}. Position-dependent masses (PDM) also hold out to deformation in the quantum canonical commutation relations or curvature of the underlying space \cite{cq04, infeld, carinena}. Furthermore, they may also appear in Hermitian Hamiltonians equivalent to $\cal PT$-symmetric or pseudo-Hermitian ones \cite{bagchi06a, bagchi06b}, whose study is a topic of current considerable interest \cite{bender}.

It is worth stressing that in addition to their position dependence responsible for noncommutativity with the momentum \cite{vonroos}, some PDM may present unusual characteristics, such as discontinuities, singularities or vanishing at one or both end points of the interval. These pro\-per\-ties may create some problems, which are not present in the constant-mass case, but are dealt with in the literature. For instance, a vanishing of the PDM, which often occurs in curved spaces \cite{cq04, infeld, carinena} with applications to quantum dots \cite{gritsev}, imposes some supplementary condition on bound-state wavefunctions \cite{bagchi05a}.

Much attention has recently been devoted to f\/inding exact solutions to Schr\"odinger equations in a PDM context due to their usefulness in physical applications. Such studies use all kinds of methods known for solving constant-mass Schr\"odinger equations or an extension of them. References on the subject can be traced, for example, through~\cite{cq08}. In the present paper, we quote only the ones that are directly relevant to what we are going to deal with.

On building on the previously noted equivalence between the presence of a PDM and a~deformation of the commutation relations \cite{cq04}, we have recently devised a method for generating pairs of potential and mass for which the Schr\"odinger equation is exactly solvable \cite{bagchi05a}. For such a purpose, we have used an approach inspired by a branch of supersymmetric quantum mechanics \cite{cooper, junker, bagchi00}, whose development dates back to that of quantum groups and $q$-algebras \cite{spiridonov92a, spiridonov92b, khare, barclay, sukhatme, gango, loutsenko}. It consists in considering those one-dimensional potentials that are translationally shape invariant for a constant mass and in deforming the corresponding shape invariance condition in such a way that it remains solvable.

This type of method has easily provided us with exact results for the bound-state energy spectrum and for the ground-state wavefunction. However the derivation of the excited-state wavefunctions in explicit form has turned out to be far more tricky because they are expressed in terms of polynomials satisfying some dif\/ferential-dif\/ference equation, whose solution is rather dif\/f\/icult in general form. Up to now we have only been able to directly solve it in the case of the $d$-dimensional radial oscillator~\cite{cq07}. This result has then been extended to the Morse and the $D$-dimensional Coulomb potentials \cite{cq08} by using a point canonical transformation (PCT) analogous to that relating the constant-mass problems \cite{bhatta, levai, de}. In doing so, we have taken advantage of the fact that the three potentials belong to the so-called Natanzon conf\/luent potential class~\mbox{\cite{natanzon, cordero}}.

The remaining point at issue is therefore the explicit form of the excited-state wavefunctions for those potentials of~\cite{bagchi05a} that do not belong to this class. It is the purpose of the present paper to answer such a question. To this end, it will prove expedient to combine the deformed shape invariance (DSI) technique with the PCT method in a novel way. Since we already know the energy spectrum of each PDM problem, we may try to f\/ind a constant-mass Schr\"odinger equation giving rise to a similar type of spectrum and to devise a PCT mapping the former problem onto the latter. It will then be a simple task to employ the reciprocal PCT to derive the PDM Schr\"odinger equation wavefunctions from the knowledge of the constant-mass ones. As we plan to show, this combined method works for all pairs of potential and mass considered in \cite{bagchi05a}, including those already solved in~\cite{cq08, cq07}.

This paper is organized as follows. In Section~\ref{section2}, the application of the DSI method to PDM Schr\"odinger equations is reviewed and the previously obtained results summarized. In Section~\ref{section3}, the use of the PCT method is explained by means of two detailed examples and the wavefunctions are listed for all the potential and PDM pairs. Finally, Section~\ref{section4} contains the conclusion.

\section{Deformed shape invariance method}\label{section2}

In one-dimensional nonrelativistic quantum mechanics, one may deform the conventional cano\-ni\-cal commutation relation $[x, p] = {\rm i}$, where $p = - {\rm i} d/dx$ and $\hbar = 1$, into \cite{cq04}
\begin{gather}
  [x, \pi] = {\rm i} f(\alphab; x), \qquad \pi = - {\rm i} \sqrt{f(\alphab; x)} \frac{d}{dx} \sqrt{f(\alphab; x)}.
  \label{eq:CR}
\end{gather}
Here we assume that the deforming function $f(\alphab; x)$ is real, positive, and depends on a set of real parameters $\alphab$, in such a way that $f(\alphab; x) \to 1$ in the $\alphab \to 0$ limit. In the following, it proves convenient to write $f(\alphab; x) = 1 + g(\alphab; x)$. Note that from (\ref{eq:CR}), it results that both $x$ and $\pi$ are Hermitian operators.

By substituting $\pi^2$ for $p^2$ in the conventional Schr\"odinger equation
\begin{gather}
  H \psi_n(x) \equiv [p^2 + V(x)] \psi_n(x) = E_n \psi_n(x),  \label{eq:SE}
\end{gather}
we arrive at a deformed equation
\begin{gather}
  H^{(\alphab)} \psi_n^{(\alphab)}(x) \equiv [\pi^2 + V(x)] \psi_n^{(\alphab)}(x) = E_n^{(\alphab)}
  \psi_n^{(\alphab)}(x).  \label{eq:def-SE}
\end{gather}
The latter may be re-interpreted as a PDM one,
\begin{gather}
  \left(- \frac{d}{dx} \frac{1}{M(\alphab; x)} \frac{d}{dx} + V_{\rm ef\/f}(\alphab; x)\right) \psi_n^{(\alphab)}(x)
  = E_n^{(\alphab)} \psi_n^{(\alphab)}(x),  \label{eq:PDM-SE}
\end{gather}
with a mass and an ef\/fective potential given by
\begin{gather}
  M(\alphab; x) = \frac{1}{f^2(\alphab; x)}, \qquad V_{\rm ef\/f}(\alphab; x) = V(x) - \frac{1}{2} f(\alphab; x)
  f''(\alphab; x) - \frac{1}{4} f^{\prime2}(\alphab; x),  \label{eq:mass}
\end{gather}
respectively \cite{cq04, bagchi05a, cq08}. Here a prime stands for derivative with respect to $x$. In (\ref{eq:SE}), we have taken units wherein the constant mass $m_0 = 1/2$, while in (\ref{eq:PDM-SE}), $M(\alphab; x)$ denotes the dimensionless part of the mass function $m(\alphab; x) = m_0 M(\alphab; x)$.

It should be noted that the ordering chosen for the noncommuting momentum and mass operators in the PDM Schr\"odinger equation (\ref{eq:PDM-SE}) is that of BenDaniel and Duke \cite{bendaniel}, for which some physical arguments have been put forward~\cite{levy}. Other choices maintaining Hermiticity of the kinetic energy operator may be taken care of by adopting the von Roos ansatz, depending on two independent ambiguity parameters~\cite{vonroos}. In such a case, the ef\/fective potential, def\/ined in~(\ref{eq:mass}), contains an additional contribution coming from those parameters (see, e.g., \cite{bagchi04} for a~discussion of this topic).

The DSI method \cite{bagchi05a} considers the Hamiltonian $H^{(\alphab)}$, def\/ined in equation (\ref{eq:def-SE}), as the f\/irst member $H_0^{(\alphab)} = H^{(\alphab)}$ of a hierarchy of Hamiltonians
\begin{gather*}
  H_i^{(\alphab)} = A^+(\alphab, \lambdab_i) A^-(\alphab, \lambdab_i) + \sum_{j=0}^i \epsilon_j, \qquad
  i = 0, 1, 2, \ldots,
\end{gather*}
where the f\/irst-order dif\/ferential operators
\begin{gather*}
  A^{\pm}(\alphab, \lambdab_i) = \mp \sqrt{f(\alphab; x)} \frac{d}{dx} \sqrt{f(\alphab; x)} + W(\lambdab_i; x)
\end{gather*}
satisfy a DSI condition
\begin{gather}
  A^-(\alphab, \lambdab_i) A^+(\alphab, \lambdab_i) = A^+(\alphab, \lambdab_{i+1}) A^-(\alphab,
  \lambdab_{i+1}) + \epsilon_{i+1}, \qquad i = 0, 1, 2, \ldots,  \label{eq:DSI}
\end{gather}
and $\epsilon_i$, $i = 0, 1, 2,\ldots$, are some real constants. Solving equation (\ref{eq:DSI}) means f\/inding a superpotential $W(\lambdab; x)$ (depending on some set of real parameters $\lambdab$), a deforming function $f(\alphab; x)$ and some constants $\lambdab_i$, $\epsilon_i$, $i = 0, 1, 2,\ldots$, with $\lambdab_0 = \lambdab$, such that
\begin{gather}
  V(x) = W^2(\lambdab; x) - f(\alphab; x) W'(\lambdab; x) + \epsilon_0  \label{eq:DSI-1}
\end{gather}
and
\begin{gather}
  W^2(\lambdab_i; x) + f(\alphab; x) W'(\lambdab_i; x)  \nonumber\\
  \qquad{}
  = W^2(\lambdab_{i+1}; x)- f(\alphab; x)
  W'(\lambdab_{i+1}; x) + \epsilon_{i+1}, \qquad i = 0, 1, 2, \ldots.  \label{eq:DSI-2}
\end{gather}

\begin{table}[h!]

\caption{Potentials, deforming functions $f(\alphab; x) = 1 + g(\alphab;x)$ and ef\/fective potentials.}\label{table1}

\vspace{1mm}

\centering
\begin{tabular}{llll}
  \hline\hline\\[-0.2cm]
  Type & $V(x)$ & $g(\alphab; x)$ & $V_{\rm ef\/f}(\alphab; x) - V(x)$\\[0.2cm]
  \hline\\[-0.2cm]
  SHO$_a$ & $\frac{1}{4} \omega^2 \left(x - \frac{2d}{\omega}\right)^2$ & $\alpha x^2 + 2\beta x$ &
       $- 2(\alpha x + \beta)^2 - \delta^2$ \\[0.2cm]
  & $- \infty < x < \infty$ & $\alpha > \beta^2 \ge 0$, $\beta \ne - \frac{2\alpha d}{\omega}$
       & $\delta = \sqrt{\alpha - \beta^2}$ \\[0.2cm]
  SHO$_b$ & $\frac{1}{4} \omega^2 \left(x - \frac{2d}{\omega}\right)^2$ & $\alpha x^2 + 2\beta x$ &
       $- 2\alpha^2 \left(x - \frac{2d}{\omega}\right)^2 - \delta^2$  \\[0.2cm]
  & $- \infty < x < \infty$ & $\alpha > \beta^2 \ge 0$, $\beta = - \frac{2\alpha d}{\omega}$
       &  $\delta = \sqrt{\alpha \left(1 - \frac{4\alpha d^2}{\omega^2}\right)}$ \\[0.2cm]
  RHO & $\frac{1}{4} \omega^2 x^2 + \frac{L(L+1)}{x^2}$ & $\alpha x^2$ & $-2 \alpha^2 x^2 - \alpha$
       \\[0.2cm]
  & $0 < x < \infty$ & $\alpha > 0$  \\[0.2cm]
  C & $- \frac{2Z}{x} + \frac{L(L+1)}{x^2}$ & $\alpha x$ & $ - \frac{\alpha^2}{4}$  \\[0.2cm]
  & $0 < x < \infty$ & $\alpha > 0$  \\[0.2cm]
  M & $B^2 e^{-2x} - B(2A+1) e^{-x}$ & $\alpha e^{-x}$ & $- \frac{3}{4} \alpha^2 e^{-2x} - \frac{1}{2} \alpha
       e^{-x}$  \\[0.2cm]
  & $- \infty < x < \infty$, $A, B > 0$ & $\alpha > 0$  \\[0.2cm]
  E$_a$ & $A(A-1) \csch^2 x - 2B \coth x$ & $\alpha e^{-x} \sinh x$ & $ \alpha e^{-2x} \left(\delta -
      \frac{3}{4} \alpha e^{-2x}\right)$  \\[0.2cm]
  & $0 < x < \infty$, $A \ge \frac{3}{2}$, $B > A^2$ & $- 2 < \alpha \ne 0$ & $\delta = 1 +
      \frac{\alpha}{2}$  \\[0.2cm]
  E$_b$ & $A(A-1) \csch^2 x - 2B \coth x$ & $\alpha e^{-x} \sinh x$ & $ -3 e^{-4x}$  \\[0.2cm]
  & $0 < x < \infty$, $A \ge \frac{3}{2}$, $B > A^2$ & $\alpha = - 2$  \\[0.2cm]
  PT & $A(A-1) \sec^2 x$ & $\alpha \sin^2 x$ & $ - \alpha - 2\alpha(\alpha-1) \sin^2 x$   \\[0.2cm]
  & $- \frac{\pi}{2} < x < \frac{\pi}{2}$, $A > 1$ & $- 1 < \alpha \ne 0$ & $\: \mbox{} + 3 \alpha^2
      \sin^4 x$  \\[0.2cm]
  S & $[A(A-1) + B^2] \sec^2 x$ & $\alpha \sin x$ & $\frac{\alpha}{2} \sin x -
      \frac{\alpha^2}{4} (1 - 3 \sin^2 x)$  \\[0.2cm]
  & $\: \mbox{} - B(2A-1) \sec x \tan x$ & $0 < |\alpha| < 1$  \\[0.2cm]
  & $- \frac{\pi}{2} < x < \frac{\pi}{2}$, $A-1 > B > 0$  \\[0.2cm]
  RM & $A(A-1) \csc^2 x + 2B \cot x$ & $ \sin x (\alpha \cos x + \beta \sin x)$ & $\left(1 + \frac{\beta}{2}
      \right) (\alpha \sin 2x$ \\[0.2cm]
  & $0 < x < \pi$, $A \ge \frac{3}{2}$ & $\sqrt{1 + \beta} > \frac{|\alpha|}{2}$, $\beta > -1$ & $\: \mbox{}
      - \beta \cos 2x)$  \\[0.2cm]
  & & & $ \: \mbox{} - \frac{3}{8}(\alpha^2 - \beta^2) \cos 4x$ \\[0.2cm]
  & & & $\: \mbox{} - \frac{3}{4} \alpha \beta \sin 4x$ \\[0.2cm]
  & & & $\: \mbox{} + \frac{1}{8}(\alpha^2 + \beta^2)$ \\[0.2cm]
  \hline \hline
\end{tabular}
\vspace{-3mm}
\end{table}

On starting from the known superpotentials of translationally shape-invariant potentials \cite{cooper}, we have been able in most cases to maintain the solvability of equations (\ref{eq:DSI-1}) and (\ref{eq:DSI-2}) for $f \ne 1$ by a procedure detailed in \cite{bagchi05a}. The resulting deforming functions and deformed superpotentials are listed in Tables~\ref{table1} and~\ref{table2} for the shifted harmonic oscillator (SHO), radial harmonic oscillator (RHO), radial Coulomb (C),  Morse (M), Eckart (E), P\"oschl--Teller (PT), Scarf I (S) and Rosen--Morse I (RM) potentials, respectively. It is worth noting that the sets of parameters $\alphab$ and $\lambdab$ contain either one element (denoted by $\alpha$ and $\lambda$) or two elements (denoted by $\lambda$, $\mu$ and $\alpha$, $\beta$) and that the radial harmonic oscillator and Coulomb potentials actually act in a $D$-dimensional space, where the angular momentum quantum number $l$ is related to the parameter $L$ through the equation $L = l + \frac{D-3}{2}$. Furthermore, the Scarf II potential is missing from the lists because no positive function $f$ on the whole real line has been found. In contrast, the Rosen--Morse~II and generalized P\"oschl--Teller potentials give rise to some acceptable functions $f$, but are omitted from Tables~\ref{table1} and~\ref{table2} for a reason explained below.

\begin{table}[h!]

\caption{Superpotentials and corresponding parameters.}\label{table2}

\vspace{1mm}

\centering
\begin{tabular}{llll}
  \hline\hline\\[-0.2cm]
  Type & $W(\lambdab, x)$ & $\lambdab$ & $\lambdab_n$ \\[0.17cm]
  \hline\\[-0.2cm]
  SHO$_{a, b}$ & $\lambda x + \mu$ & $\lambda = \frac{1}{2}(\alpha + \Delta)$, $\Delta = \sqrt{\omega^2
       + \alpha^2}$ & $\lambda_n = \lambda + n\alpha$ \\[0.17cm]
  & & $\mu = \beta - \frac{d\omega}{2\lambda}$ & $ \mu_n  = \frac{\lambda \mu + 2n\beta \lambda +
       n^2 \alpha \beta}{\lambda + n\alpha}$  \\[0.17cm]
  RHO & $\frac{\lambda}{x} + \mu x$ & $\lambda = - L - 1$ & $\lambda_n = \lambda - n$  \\[0.17cm]
  & & $\mu = \frac{1}{2} (\alpha + \Delta)$, $\Delta = \sqrt{\omega^2 + \alpha^2}$ & $\mu_n = \mu +
       n\alpha$  \\[0.17cm]
  C & $\frac{\lambda}{x} + \mu$ & $\lambda = - L - 1$ & $\lambda_n = \lambda - n$  \\[0.17cm]
  & & $\mu = - \frac{2Z + \alpha\lambda}{2\lambda}$ & $ \mu_n = - \frac{2Z + \alpha \lambda (2n+1)
       - \alpha n^2}{2(\lambda - n)}$  \\[0.17cm]
  M & $\lambda e^{-x} + \mu$ & $\lambda = - \frac{1}{2} (\alpha + \Delta)$, $\Delta = \sqrt{4B^2 +
       \alpha^2}$ & $ \lambda_n = \lambda - n\alpha$  \\[0.17cm]
  & & $\mu = - \frac{1}{2} \left(\frac{B(2A+1)}{\lambda} + 1\right)$ & $\mu_n = \frac{2\lambda(\mu - n)
       + n^2 \alpha}{2(\lambda - n\alpha)}$  \\[0.17cm]
  E$_{a,b}$ & $\lambda \coth x + \mu$ & $\lambda = - A$ & $\lambda_n = \lambda - n$  \\[0.17cm]
  & & $\mu = \frac{B}{A} - \frac{1}{2} \alpha$ & $\mu_n = \frac{\lambda\mu - \frac{1}{2}\alpha n
       (2\lambda - n)}{\lambda - n}$  \\[0.17cm]
  PT & $\lambda \tan x$ & $\lambda = \frac{1}{2}(1 + \alpha + \Delta)$ & $\lambda_n = \lambda + n(1+
       \alpha)$  \\[0.17cm]
  & & $\Delta = \sqrt{(1+\alpha)^2 + 4A(A-1)}$  \\[0.17cm]
  S & $\lambda \tan x + \mu \sec x$ & $\lambda = \frac{1}{2}(1 + \Delta_+ + \Delta_-)$ & $\lambda_n =
       \lambda + n$  \\[0.17cm]
  & & $\mu = \frac{1}{2}(\alpha - \Delta_+ + \Delta_-)$ & $\mu_n = \mu + n \alpha$  \\[0.17cm]
  & & $\Delta_{\pm} = \sqrt{\frac{1}{4}(1 \mp \alpha)^2 + C_{\pm}(C_{\pm} -1)}$ \\[0.17cm]
  & & $C_{\pm} = A \pm B$  \\[0.17cm]
  RM & $\lambda \cot x + \mu$ & $\lambda = - A$ & $\lambda_n = \lambda - n$  \\[0.17cm]
  & & $\mu = - \frac{B}{A} - \frac{1}{2}\alpha$ & $\mu_n = \frac{\lambda\mu - \frac{1}{2}\alpha n
       (2\lambda - n)}{\lambda - n}$  \\[0.17cm]
  \hline \hline
\end{tabular}
\vspace{-3mm}
\end{table}

From the constants $\epsilon_i$, the energy eigenvalues are determined through the relation
\begin{gather}
  E_n^{(\alphab)} = \sum_{i=0}^n \epsilon_i.  \label{eq:E}
\end{gather}
The ground-state wavefunction, annihilated by $A^-(\alphab, \lambdab)$, can be written as
\begin{gather}
  \psi_0^{(\alphab)}(x) = \psi_0^{(\alphab)}(\lambdab; x) \propto \frac{1}{\sqrt{f(\alphab; x)}} \exp\left(-
  \int^x\frac{W(\lambdab; x')}{f(\alphab; x')} dx'\right)  \label{eq:gswf}
\end{gather}
and the excited-state wavefunctions can in principle be obtained recursively by acting with $A^+(\alphab, \lambdab)$,
\begin{gather}
  \psi_{n+1}^{(\alphab)}(x) = \psi_{n+1}^{(\alphab)}(\lambdab; x) \propto A^+(\alphab, \lambdab)
  \psi_n^{(\alphab)}(\lambdab_1; x).  \label{eq:eswf}
\end{gather}

Equation (\ref{eq:E}) only provides solutions to the bound-state energies of $H^{(\alphab)}$ if the corresponding wavefunctions (\ref{eq:gswf}) and (\ref{eq:eswf}) are physically acceptable. As stressed in \cite{bagchi05a}, this imposes not only that they are square integrable on the (f\/inite or inf\/inite) interval of def\/inition $(x_1, x_2)$ of~$V(x)$, as for the conventional Schr\"odinger equation, but also that they ensure the Hermiticity of $H^{(\alphab)}$ (or that of $\pi$). The latter condition translates into
\begin{gather*}
  \big|\psi_n^{(\alphab)}(\lambdab; x)\big|^2 f(\alphab; x) = \frac{\big|\psi_n^{(\alphab)}(\lambdab; x)
  \big|^2}{\sqrt{M(\alphab; x)}} \to 0 \qquad \text{for} \quad x \to x_1 \quad \text{and} \quad x \to x_2,
\end{gather*}
which imposes an additional restriction whenever $f(\alphab; x) \to \infty$ (or $M(\alphab; x) \to 0$) for $x \to x_1$ and/or $x \to x_2$. It is actually this condition that is not satisf\/ied by the square-integrable wavefunctions obtained for the Rosen--Morse II and generalized P\"oschl--Teller potentials and explains their absence from Tables~\ref{table1} and~\ref{table2}.

The resulting bound-state energies $E_n^{(\alphab)}$, corresponding to the potentials and deforming functions of Table~\ref{table1}, are listed in Table~\ref{table3}. This table illustrates the strong inf\/luence that the deformation or mass parameters may have on the spectrum. This is particularly striking for the Coulomb potential, whose inf\/inite number of bound states for a constant mass is converted into a f\/inite one, and for the Eckart potential, for which a f\/inite number of bound states becomes inf\/inite in the case $\alpha = -2$. We plan to come back to these features in Section~\ref{section3} and to provide there a simple derivation of the bound-state number.

\begin{table}[h!]

\caption{Bound-state energy spectra.}\label{table3}
\vspace{1mm}

{

\centering
\begin{tabular}{lll}
  \hline\hline\\[-0.2cm]
  Type & $E^{(\alphab)}_n$ & $n$\\[0.17cm]
  \hline\\[-0.2cm]
  SHO$_{a, b}$ & $(2n+1) \lambda + n^2 \alpha +d^2 - \left(\frac{\lambda\mu + 2n\beta\lambda + n^2
       \alpha\beta}{\lambda + n\alpha}\right)^2$ & $0, 1, 2, \ldots$  \\[0.17cm]
  RHO & $- 2\lambda\mu - \alpha\lambda + \mu - 4(\alpha\lambda - \mu) n + 4\alpha n^2$ & $0, 1, 2,
       \ldots$  \\[0.17cm]
  C & $- \frac{1}{4} \left(\frac{2Z + \alpha\lambda(2n+1) - \alpha n^2}{\lambda - n}\right)^2$ & $0, 1, \ldots,
       n_{\rm max}$  \\[0.17cm]
  & & $n_{\rm max}^2 + |\lambda|(2n_{\rm max} + 1) < \frac{2Z}{\alpha}$ \\[0.17cm]
  & & $\: \mbox{} \le (n_{\rm max} + 1)^2 + |\lambda|(2n_{\rm max} + 3)$  \\[0.17cm]
  M & $- \frac{1}{4}\left(\frac{2\lambda(\mu-n) + n^2 \alpha}{\lambda - n\alpha}\right)^2$ & $0, 1, \ldots,
       n_{\rm max}$  \\[0.17cm]
  & & $n_{\rm max}(2|\lambda| + n_{\max} \alpha) < 2|\lambda|\mu$  \\[0.17cm]
  & & $\: \mbox{} \le (n_{\rm max} + 1)[2|\lambda| + (n_{\max} + 1) \alpha]$  \\[0.17cm]
  E$_a$ & $- (\lambda - n)^2 - \left(\frac{2\lambda\mu - \alpha n(2\lambda - n)}{2(\lambda - n)}\right)^2$
       & $0, 1, \ldots, n_{\rm max}$  \\[0.17cm]
  & $\: \mbox{} + \alpha [(2n+1)\lambda - n^2]$ & $(|\lambda| + n_{\rm max})^2 < \frac{|\lambda| (2\mu
       + \alpha|\lambda|)}{2\delta}$  \\[0.17cm]
  & & $\: \mbox{} \le (|\lambda| + n_{\rm max} + 1)^2$  \\[0.17cm]
  E$_b$ & $- (\lambda - n)^2 - \left(\frac{\lambda\mu + n(2\lambda - n)}{\lambda - n}\right)^2$
       & $0, 1, 2, \ldots$  \\[0.17cm]
  & $\: \mbox{} - 2 [(2n+1)\lambda - n^2]$ \\[0.17cm]
  PT & $(\lambda + n)^2 - \alpha (\lambda - n^2)$ &  $0, 1, 2, \ldots$  \\[0.17cm]
  S$^a$ & $(\lambda + n)^2 - \alpha (2n+1) \mu - \alpha^2 n^2$ &  $0, 1, 2, \ldots$  \\[0.17cm]
  RM & $(\lambda - n)^2 - \left(\frac{2\lambda\mu - \alpha n(2\lambda - n)}{2(\lambda - n)}\right)^2$
       & $0, 1, 2, \ldots$  \\[0.17cm]
  & $\: \mbox{} - \beta (2n+1) \lambda + \beta n^2$  \\[0.17cm]
  \hline \hline
\end{tabular}

}
\vspace{1mm}

\noindent
$^a$ A misprint has been corrected in the Appendix of \cite{bagchi05a}.
\end{table}

The corresponding wavefunctions are listed in Table~\ref{table4}, where they are given in terms of $n$th-degree polynomials $P_n(\lambdab; y)$ in some new variable $y$, such that $P_0(\lambdab; y) \equiv 1$. As a consequence of the recursion relation (\ref{eq:eswf}), those polynomials satisfy dif\/ferential-dif\/ference equations, given in Table~\ref{table5}. From such equations, it is clear that if the ground-state and f\/irst few excited-state wavefunctions are easy to obtain, the same is not true for the remaining ones.

\begin{table}[h!]

\caption{Bound-state wavefunctions resulting from the DSI approach.}\label{table4}

\vspace{1mm}

\centering
\begin{tabular}{lll}
  \hline\hline\\[-0.2cm]
  Type & $\psi^{(\alphab)}_n(\lambdab; x)$ & $y$\\[0.2cm]
  \hline\\[-0.2cm]
  SHO$_{a, b}$ & $f^{- \frac{(\lambda_n + \alpha)}{2\alpha}} \exp \left(- \frac{\beta\lambda_n - \alpha
       \mu_n}{\alpha\delta} \arccot\frac{\alpha x + \beta}{\delta}\right) P_n(\lambda, \mu; y)$ & $x$
       \\[0.2cm]
  RHO & $x^{|\lambda|} f^{- \frac{\mu_n + (|\lambda_n|+1) \alpha}{2\alpha}} P_n(\lambda, \mu; y)$ & $x^2$
       \\[0.2cm]
  C & $x^{|\lambda_n|} f^{- \frac{2\mu_n + (2|\lambda_n|+1) \alpha}{2\alpha}} P_n(\lambda, \mu; y)$ &
       $x^{-1}$  \\[0.2cm]
  M & $f^{\frac{2\lambda_n - (2\mu_n + 1) \alpha}{2\alpha}} e^{- \mu_n x} P_n(\lambda, \mu; y)$ &
       $e^{-x}$  \\[0.2cm]
  E$_a$ & $(\coth x + 1)^{\frac{1}{2}} (\coth x + 1 + \alpha)^{\frac{(1+\alpha)\lambda_n - \mu_n}{2\delta}
       - \frac{1}{2}} (\coth x - 1)^{\frac{\lambda_n + \mu_n}{2\delta}}  P_n(\lambda, \mu; y)$ & $\coth x$
       \\[0.2cm]
  E$_b$ & $ (\coth x - 1)^{\lambda_n - 1} \csch x \exp\left(- \frac{\lambda_n + \mu_n}{\coth x - 1}\right)
       P_n(\lambda, \mu; y)$ & $\coth x$  \\[0.2cm]
  PT & $(\cos x)^{\frac{\lambda_n}{1+\alpha}} f^{- \frac{\lambda_n + 1 + \alpha}{2(1 + \alpha)}}
       P_n(\lambda; y)$ & $\tan x$  \\[0.2cm]
  S & $f^{- \frac{\lambda_n - \alpha \mu_n}{1 - \alpha^2} - \frac{1}{2}} (1 - \sin x)^{\frac{\lambda+\mu}
       {2(1 + \alpha)}} (1 + \sin x)^{\frac{\lambda-\mu}{2(1 - \alpha)}} P_n(\lambda, \mu; y)$ & $\sin x$
       \\[0.2cm]
  RM & $f^{- \frac{1}{2}(|\lambda_n| + 1)} (\sin x)^{|\lambda_n|} \exp\left(- \frac{2\mu_n + \alpha|\lambda_n|}
       {2\delta} \arccot \frac{\cot x + \frac{\alpha}{2}}{\delta}\right) P_n(\lambda, \mu; y)$ & $\cot x$
       \\[0.2cm]
  & $\delta = \sqrt{1 + \beta - \frac{\alpha^2}{4}}$  \\[0.2cm]
  \hline\hline
\end{tabular}
\end{table}

\begin{table}[h!]

\caption{Dif\/ferential-dif\/ference equations satisf\/ied by the polynomials.}\label{table5}

\vspace{1mm}

\centering
\begin{tabular}{ll}
  \hline\\[-0.2cm]
  Type & Equation for $P_n(\lambdab; y)$\\[0.2cm]
  \hline\hline\\[-0.2cm]
  SHO$_{a, b}$ & $P_{n+1}(\lambda, \mu; y) = \left(- (1 + 2\beta y + \alpha y^2) \frac{d}{dy} +
       (\lambda_{n+1} + \lambda) y + \mu_{n+1} + \mu\right) P_n(\lambda_1, \mu_1; y)$  \\[0.2cm]
  RHO & $P_{n+1}(\lambda, \mu; y) = \left(- 2y (1 + \alpha y) \frac{d}{dy} + \lambda_{n+1} + \lambda + n
       + (\mu_{n+1} + \mu + n\alpha) y\right)$  \\[0.2cm]
  & $\: \mbox{} \times  P_n(\lambda_1, \mu_1; y)$  \\[0.2cm]
  C & $P_{n+1}(\lambda, \mu; y) = \left(y (y + \alpha) \frac{d}{dy} + (\lambda_{n+1} + \lambda) y
       + \mu_{n+1} + \mu\right) P_n(\lambda_1, \mu_1; y)$  \\[0.2cm]
  M & $P_{n+1}(\lambda, \mu; y) = \left(y (1 + \alpha y) \frac{d}{dy} + (\lambda_{n+1} + \lambda) y
       + \mu_{n+1} + \mu\right) P_n(\lambda_1, \mu_1; y)$  \\[0.2cm]
  E$_{a,b}$ & $P_{n+1}(\lambda, \mu; y) = \left((y^2 + \alpha y - 1 - \alpha) \frac{d}{dy} +
       (\lambda_{n+1} + \lambda) y + \mu_{n+1} + \mu\right) P_n(\lambda_1, \mu_1; y)$  \\[0.2cm]
  PT & $P_{n+1}(\lambda; y) = \left(- [1 + (1 + \alpha)y^2] \frac{d}{dy} + (\lambda_{n+1} + \lambda) y
       \right) P_n(\lambda_1; y)$  \\[0.2cm]
  S & $P_{n+1}(\lambda, \mu; y) = \left(- (1 + \alpha y)(1 - y^2) \frac{d}{dy} - ny(1 + \alpha y)
       + (\lambda_{n+1} + \lambda) y + \mu_{n+1} + \mu\right)$  \\[0.2cm]
  & $\: \mbox{} \times  P_n(\lambda_1, \mu_1; y)$  \\[0.2cm]
  RM & $P_{n+1}(\lambda, \mu; y) = \left((y^2 + \alpha y + 1 + \beta) \frac{d}{dy} +
       (\lambda_{n+1} + \lambda) y + \mu_{n+1} + \mu\right) P_n(\lambda_1, \mu_1; y)$  \\[0.2cm]
  \hline\hline
\end{tabular}
\end{table}

To obtain all the bound-state wavefunctions in explicit form, we shall proceed in Section~\ref{section3} to combine the DSI approach with the PCT method.

\section{Point canonical transformation method}\label{section3}

\subsection{General method}

As shown in \cite{bagchi04}, a constant-mass Schr\"odinger equation
\begin{gather}
  \left(- \frac{d^2}{du^2} + U(u)\right) \phi_n(u) = \varepsilon_n \phi_n(u),  \label{eq:SE-bis}
\end{gather}
for some potential $U(u)$ def\/ined on a f\/inite or inf\/inite interval, can be transformed into a PDM one, given in equation
(\ref{eq:PDM-SE}), via a change of variable
\begin{gather}
  u(\alphab; x) = a v(\alphab; x) + b, \qquad v(\alphab; x) = \int^x \sqrt{M(\alphab; x')}\, dx',
  \label{eq:variable}
\end{gather}
and a change of function
\begin{gather}
  \phi_n(u(\alphab; x)) \propto [M(\alphab; x)]^{-1/4} \psi_n^{(\alphab)}(x).  \label{eq:function}
\end{gather}
In (\ref{eq:variable}), $a$ and $b$ are assumed real. The ef\/fective potential, def\/ined on a possibly dif\/ferent interval, and the energy eigenvalues of the PDM Schr\"odinger equation are given in terms of the potential and the energy eigenvalues of the constant-mass one by
\begin{gather}
  V_{\rm ef\/f}(\alphab; x) = a^2 U(a v(\alphab; x) + b) + \frac{M''}{4M^2} - \frac{7M^{\prime2}}{16M^3} + c
  \label{eq:potential}
\end{gather}
and
\begin{gather}
  E_n^{(\alphab)} = a^2 \varepsilon_n + c,  \label{eq:energy}
\end{gather}
where, as before, a prime denotes derivative with respect to $x$ and we have introduced an additive real constant $c$, not present in~\cite{bagchi04}.

We can reformulate the PCT, def\/ined in (\ref{eq:variable}) and (\ref{eq:function}), in terms of the deforming function $f(\alphab; x)$ as
\begin{gather}
  u(\alphab; x)  = a v(\alphab; x) + b, \qquad v(\alphab; x) = \int^x \frac{dx'}{f(\alphab; x')},
      \label{eq:variable-bis} \\
  \phi_n(u(\alphab; x))   \propto \sqrt{f(\alphab; x)}\, \psi_n^{(\alphab)}(x).  \label{eq:function-bis}
\end{gather}
Then, on taking (\ref{eq:mass}) into account, equation (\ref{eq:potential}) is replaced by
\begin{gather}
  V(x) = a^2 U(a v(\alphab; x) + b) + c.  \label{eq:potential-bis}
\end{gather}

In most applications, this PCT is used in the following way (see, e.g., \cite{alhaidari}). One starts from a given exactly solvable constant-mass Schr\"odinger equation, hence from some known~$U(u)$,~$\varepsilon_n$, and $\phi_n(u)$. One chooses a PDM and some parameters $a$, $b$, which means some change of va\-riab\-le~(\ref{eq:variable}). As a result, one obtains an exactly solvable PDM Schr\"odinger equation, containing an ef\/fective potential given by (\ref{eq:potential}), and whose eigenvalues and eigenfunctions can be directly derived from equations (\ref{eq:energy}) and (\ref{eq:function}), respectively. There are variants of this approach, wherein the dependence of the variable $u$ on the PDM is dif\/ferent from that given in (\ref{eq:variable}), but otherwise the method remains the same (for some examples see, e.g.,~\cite{bagchi05b}).

Here, in contrast, we are going to make use of our acquaintance with $M(\alphab; x)$ (or $f(\alphab; x)$), $V_{\rm ef\/f}(\alphab; x)$ (or $V(x)$), and $E_n^{(\alphab)}$, coming from the DSI method, to determine $\psi_n^{(\alphab)}(x)$. From the dependence of $E_n^{(\alphab)}$ on $n$, we f\/irst guess which type of potential $U(u)$ in the constant-mass Schr\"odinger equation (\ref{eq:SE-bis}) may give rise to such a dependence through equation (\ref{eq:energy}). Furthermore, from the latter equation, we determine the values of the constants $a^2$ and $c$, as well as all or only some parameters of~$U(u)$.\footnote{It may happen for some potentials $U(u)$ that not all their parameters appear in their eigenvalues $\varepsilon_n$. For the Scarf I potential, for instance, $\varepsilon_n$ only depends on $A$, but not on $B$. In such cases, equation (\ref{eq:energy}) is not enough to entirely determine~$U(u)$.} In a second step, we obtain the change of variable~(\ref{eq:variable-bis}) from our knowledge of $f(\alphab; x)$ and $|a|$, the sign of $a$ and the value of $b$ being determined from the known domain of def\/inition of $U(u)$. The third step then consists in checking equation (\ref{eq:potential-bis}) and calculating the remaining unknown parameters of $U(u)$, whenever there are some. Finally, since the wavefunctions $\phi_n(u)$ of equation (\ref{eq:SE-bis}) are explicitly known, we deduce $\psi_n^{(\alphab)}(x)$ from equation~(\ref{eq:function-bis}).

Before listing the results for $\psi_n^{(\alphab)}(x)$, we shall proceed to illustrate our method with two examples: a simple one (corresponding to the P\"oschl--Teller potential), wherein the PCT does not change the nature of the potential, and a more complicated one (corresponding to the Morse potential), wherein the PCT deeply modif\/ies the potential and the number of bound states acquires an intricate dependence on the parameters.

\subsection[P\"oschl-Teller potential]{P\"oschl--Teller potential}

From Tables \ref{table1} and~\ref{table3}, we observe that for the deforming function $f(\alpha; x) = 1 + \alpha \sin^2x$ ($- 1 < \alpha \ne 0$), the P\"oschl--Teller potential $V(x) =A(A-1) \sec^2 x$, def\/ined on $(- \frac{\pi}{2}, \frac{\pi}{2})$, has a quadratic bound-state energy spectrum made of an inf\/inite number of levels,
\begin{gather}
  E_n^{(\alpha)} = (\lambda + n)^2 - \alpha \left(\lambda - n^2\right), \qquad n = 0, 1, 2, \ldots.  \label{eq:PT-E}
\end{gather}
On the other hand, it is well known \cite{cq99} that the same type of potential
\begin{gather}
  U(u) = A'(A'-1) \sec^2 u,  \label{eq:PT-potential}
\end{gather}
def\/ined on $(- \frac{\pi}{2}, \frac{\pi}{2})$ and used with a constant mass, has a spectrum
\begin{gather}
  \varepsilon_n = (A' + n)^2, \qquad n = 0, 1, 2, \ldots,  \label{eq:PT-E-bis}
\end{gather}
with essentially the same characteristics. Equations (\ref{eq:energy}), (\ref{eq:PT-E}), and (\ref{eq:PT-E-bis}) then lead to the relations
\begin{gather}
  a^2 = 1 + \alpha, \qquad A' = \frac{\lambda}{1+\alpha}, \qquad c = \frac{\alpha}{1+\alpha} \lambda
  (\lambda - 1 - \alpha) = \frac{\alpha}{1+\alpha} A(A-1),  \label{eq:PT-para}
\end{gather}
where in the last step we used the expression of $\lambda$ given in Table~\ref{table2}.

The second equation in (\ref{eq:variable-bis}) now yields\footnote{Here and in other cases, we use the principal value of all inverse trigonometric functions.}
\begin{gather*}
  v(\alpha; x) = \frac{1}{\sqrt{1+\alpha}} \arctan\left(\sqrt{1+\alpha} \tan x\right).
\end{gather*}
Furthermore, it is obvious that if we assume $a = \sqrt{1+\alpha}$ and $b = 0$ in the f\/irst one, the variable
\begin{gather}
  u(\alpha; x) = \arctan\left(\sqrt{1+\alpha} \tan x\right)  \label{eq:PT-ux}
\end{gather}
is def\/ined on $(- \frac{\pi}{2}, \frac{\pi}{2})$, as it should be. From the inverse transformation, we get
\begin{gather}
  \tan x = \frac{\tan u}{1+\alpha}.  \label{eq:PT-xu}
\end{gather}

On inserting (\ref{eq:PT-xu}) in the def\/inition of the starting P\"oschl--Teller potential, we arrive at the expression
\begin{gather*}
  V(x) = \frac{A(A-1)}{1+\alpha} \left(\sec^2 u + \alpha\right),
\end{gather*}
which agrees with equation (\ref{eq:potential-bis}) when we take equations (\ref{eq:PT-potential}) and (\ref{eq:PT-para}) into account.

Since, for constant mass, the bound-state P\"oschl--Teller wavefunctions can be expressed in terms of Gegenbauer polynomials as \cite{cq99}
\begin{gather*}
  \phi_n(u) \propto (\cos u)^{A'} C_n^{(A')}(\sin u), \qquad n = 0, 1, 2, \ldots,
\end{gather*}
the use of equations (\ref{eq:function-bis}) and (\ref{eq:PT-ux}) leads to the searched for wavefunctions
\begin{gather*}
  \psi_n^{(\alpha)}(x) \propto [f(\alpha; x)]^{- \frac{1}{2} (A'+1)} (\cos x)^{A'} C_n^{(A')}(t), \qquad
        n = 0, 1, 2, \ldots, \\
  t = \sqrt{\frac{1+\alpha}{f(\alpha; x)}}, \qquad f(\alpha; x) = 1 + \alpha \sin^2 x,
\end{gather*}
for the P\"oschl--Teller potential and the present PDM. Here $A'$ should be replaced by its expression in terms of $A$ and $\alpha$, given in equation (\ref{eq:PT-para}) and in Table~\ref{table2}.

As a f\/inal check, we may compare the obtained wavefunctions with those coming from the DSI method and given in Tables~\ref{table4} and~\ref{table5}. Knowing the former indeed provides us with a hint for solving the dif\/ferential-dif\/ference equation satisf\/ied by the polynomials $P_n(\lambda; y)$. The changes of variable
\begin{gather*}
  t = y \sqrt{\frac{1+\alpha}{1 + (1+\alpha)y^2}} \qquad \text{or} \qquad y = \frac{t}{\sqrt{(1+\alpha)
  (1-t^2)}}
\end{gather*}
and of function
\begin{gather}
  P_n(\lambda; y) = \gamma_n^{(A')} \left(1-t^2\right)^{-n/2} C_n^{(A')}(t),  \label{eq:PT-P}
\end{gather}
where $\gamma_n^{(A')}$ is some constant, transform the equation fulf\/illed by $P_n$ into the backward shift operator relation for Gegenbauer polynomials~\cite{koekoek},
\begin{gather}
  - \frac{(n+1)(2\alpha+n-1)}{2(\alpha-1)} C_{n+1}^{(\alpha-1)}(t) = \left(\left(1-t^2\right) \frac{d}{dt} + (1-2\alpha) t
  \right) C_n^{(\alpha)}(t),  \label{eq:gegenbauer}
\end{gather}
where $\alpha$ is replaced by $A'$. In (\ref{eq:PT-P}), we have to choose
\begin{gather*}
  \gamma_n^{(A')} = (1+\alpha)^{n/2} \frac{n!\, \Gamma(2A'+2n) \Gamma (A')}{2^n \Gamma(2A'+n)
  \Gamma(A'+n)}.
\end{gather*}
This completes the derivation of wavefunctions for the P\"oschl--Teller potential.

\subsection{Morse potential}

From Tables~\ref{table1}, \ref{table2}, and~\ref{table3}, we observe that for the Morse potential $V(x) = B^2 e^{-2x} - B(2A+1) e^{-x}$ ($A$, $B > 0$) and the deforming function $f(\alpha; x) = 1 + \alpha e^{-x}$ ($\alpha > 0$), def\/ined on the real line $- \infty < x < \infty$, the bound-state energy spectrum is made of a f\/inite number of levels, whose energies can be written as
\begin{gather}
  E_n^{(\alpha)} = - \mu_n^2 = - \frac{1}{4\alpha^2} \left(\lambda_n - \frac{B[B + \alpha(2A+1)]}{\lambda_n}
  \right)^2, \qquad n = 0, 1, \ldots, n_{\max},  \label{eq:M-E}
\end{gather}
where $\lambda_n = \lambda - n\alpha$ and $n_{\max}$ is such that
\begin{gather}
  n_{\max} (2|\lambda| + n_{\max}\alpha) < 2|\lambda|\mu \le (n_{\max} + 1) [2|\lambda| +
  (n_{\max} + 1)\alpha].  \label{eq:M-condition}
\end{gather}
Such an energy spectrum looks like that of an Eckart potential
\begin{gather*}
  U(u) = A'(A'-1) \csch^2 u - 2B' \coth u, \qquad 0 < u < \infty,
\end{gather*}
for a constant mass, which is given by \cite{cooper}
\begin{gather}
   \varepsilon_n = - (A'+n)^2 - \left(\frac{B'}{A'+n}\right)^2, \qquad n = 0, 1, \ldots, n'_{\max},
       \label{eq:M-E-bis} \\
   (A' + n'_{\max})^2 < B' \le (A' + n'_{\max} + 1)^2.  \label{eq:M-condition-bis}
\end{gather}
By comparing (\ref{eq:M-E}) and (\ref{eq:M-E-bis}) with (\ref{eq:energy}), it is indeed straightforward to obtain
\begin{gather}
  a^2 = \frac{1}{4}, \qquad A' = \frac{|\lambda|}{\alpha}, \qquad B' = \frac{1}{\alpha^2} B [B + \alpha(2A+1)],
  \qquad c = \frac{1}{2} B'.  \label{eq:M-para}
\end{gather}
Furthermore, with such parameters $A'$ and $B'$, condition (\ref{eq:M-condition-bis}) directly leads to restriction (\ref{eq:M-condition}) if we set $n'_{\max} = n_{\max}$. Hence, the latter, whose origin in \cite{bagchi05a} was based on the behaviour of the wavefunctions $\psi_n^{(\alpha)}(x)$ for $x \to \pm \infty$, gets here a very simple derivation in terms of the PCT.

The remaining steps of the procedure work as for the P\"oschl--Teller potential. We successively obtain
\begin{gather*}
  v(\alpha; x) = \ln(e^x + \alpha), \qquad a = \tfrac{1}{2}, \qquad b = - \tfrac{1}{2} \ln \alpha,
\end{gather*}
leading to
\begin{gather*}
  0 < u(\alpha; x) = \frac{1}{2} \ln \frac{e^x + \alpha}{\alpha} < \infty,
\end{gather*}
and
\begin{gather}
  \psi_n^{(\alpha)}(x)  \propto f^{- \frac{1}{2} \left(A'+n+1 + \frac{B'}{A'+n}\right)} \exp\left[\frac{1}{2}
      \left(A'+n - \frac{B'}{A'+n}\right) x\right] \nonumber\\
 \phantom{\psi_n^{(\alpha)}(x)  \propto}{} \times P_n^{\left(-A'-n + \frac{B'}{A'+n}, -A'-n - \frac{B'}{A'+n}\right)}(t), \qquad t = 1 + 2\alpha
      e^{-x},  \label{eq:M-wf}
\end{gather}
where in the last step we used the known wavefunctions $\phi_n(u)$ of the Eckart potential for constant mass in terms of Jacobi polynomials \cite{cooper}. In (\ref{eq:M-wf}), $A'$ and $B'$ should be expressed in terms of~$A$,~$B$, and~$\alpha$, as shown in (\ref{eq:M-para}) and Table~\ref{table2}.

In the present case, the dif\/ferential-dif\/ference equation for $P_n(\lambda, \mu; y)$, given in Table~\ref{table5}, amounts to a combination of the recursion and dif\/ferential relations for Jacobi polynomials (see equations (22.7.1) and (22.8.1) of~\cite{abramowitz}),
\begin{gather}
  \frac{2(n+1)(n+\alpha+\beta+1)}{2n+\alpha+\beta+2} P_{n+1}^{(\alpha, \beta)}(t) \nonumber\\
\qquad  {} = \left[- (1-t^2) \frac{d}{dt} + (n+\alpha+\beta+1) \left(t + \frac{\alpha-\beta}
       {2n+\alpha+\beta+2}\right)\right] P_n^{(\alpha, \beta)}(t),  \label{eq:jacobi-2}
\end{gather}
with $\alpha$ and $\beta$ replaced by $-A'-n-1 + \frac{B'}{A'+n+1}$ and $-A'-n-1 - \frac{B'}{A'+n+1}$, respectively. The relation between both approaches is obtained through the transformation
\begin{gather*}
  P_n(\lambda, \mu; y)   = \gamma_n^{(A')} P_n^{\left(-A'-n + \frac{B'}{A'+n}, -A'-n - \frac{B'}{A'+n}\right)}
       (t), \qquad t = 1 + 2\alpha y, \\
  \gamma_n^{(A')}  = \frac{n!\, \Gamma(2A'+2n) \Gamma(A')}{2^n \Gamma(2A'+n) \Gamma(A'+n)}.
\end{gather*}

It is worth observing that the expression (\ref{eq:M-wf}), obtained here for the bound-state wavefunctions $\psi_n^{(\alpha)}(x)$, dif\/fers from equation~(3.16) of~\cite{cq08}, which is the result of a PCT starting from the PDM radial oscillator Schr\"odinger equation. In our previous result, the Jacobi polynomials indeed depend on the variable $(-1 + \alpha e^{-x})/(1 + \alpha e^{-x})$ instead of $t = 1 + 2\alpha e^{-x}$, while the other factors are also slightly modif\/ied. On writing the Jacobi polynomials in terms of hypergeometric functions and using a linear transformation formula for the latter (see equations~(15.4.6) and~(15.3.4) of \cite{abramowitz}), it is however straightforward to check that both results are in fact equivalent, as it should be.

\subsection{Results}

For the potentials and deforming functions of Table~\ref{table1}, we list the changes of variable (\ref{eq:variable-bis}), the potentials $U(u)$ used in (\ref{eq:SE-bis}), and the resulting PDM wavefunctions $\psi_n^{(\alphab)}(x)$ in Tables~\ref{table6},~\ref{table7}, and~\ref{table8}, respectively.

\begin{table}[h!]

\caption{Variables $u(\alphab; x) = a v(\alphab; x) + b$ and their domain of def\/inition.}\label{table6}

\vspace{1mm}

\centering
\begin{tabular}{lllll}
  \hline\hline\\[-0.2cm]
  Type & $v(\alphab; x)$ & $a$ & $b$ & Domain\\[0.2cm]
  \hline\\[-0.2cm]
  SHO$_a$ & $- \frac{1}{\delta} \arccot \frac{\alpha x + \beta}{\delta}$ & $- \delta$ & $0$ & $0 < u < \pi$
       \\[0.2cm]
  SHO$_b$ & $- \frac{1}{\delta} \arccot\left[\frac{\alpha}{\delta} \left(x - \frac{2d}{\omega}\right)\right]$ &
       $\delta$ & $\frac{\pi}{2}$ & $- \frac{\pi}{2} < u < \frac{\pi}{2}$  \\[0.2cm]
  RHO & $\frac{1}{\sqrt{\alpha}} \arctan(\sqrt{\alpha} x)$ & $2 \sqrt{\alpha}$ & $- \frac{\pi}{2}$ &
       $- \frac{\pi}{2} < u < \frac{\pi}{2}$  \\[0.2cm]
  C & $\frac{1}{\alpha} \ln(1 + \alpha x)$ & $\frac{\alpha}{2}$ & $0$ & $0 < u < \infty$  \\[0.2cm]
  M & $\ln(e^x + \alpha)$ & $\frac{1}{2}$ & $ - \frac{1}{2} \ln\alpha$ & $0 < u < \infty$  \\[0.2cm]
  E$_a$ & $\frac{1}{\delta} \arccoth \frac{\coth x + \frac{\alpha}{2}}{\delta}$ & $\delta$ & $0$ & $0 < u <
       \infty$  \\[0.2cm]
  E$_b$ & $(\coth x - 1)^{-1}$ & $1$ & $0$ & $0 < u < \infty$  \\[0.2cm]
  PT & $\frac{1}{\sqrt{1 + \alpha}} \arctan\left(\sqrt{1 + \alpha} \tan x\right)$ & $\sqrt{1 + \alpha}$ & $0$
       & $- \frac{\pi}{2} < u < \frac{\pi}{2}$  \\[0.2cm]
  S & $\frac{2}{\sqrt{1 - \alpha^2}} \arctan \frac{\tan\frac{x}{2} + \alpha}{\sqrt{1 - \alpha^2}}$ &
       $\sqrt{1 - \alpha^2}$ & $- \frac{\pi}{2} + 2 \arctan \sqrt{\frac{1 - \alpha}{1 + \alpha}}$ & $- \frac{\pi}
       {2} < u < \frac{\pi}{2}$  \\[0.2cm]
  RM & $\frac{1}{\delta} \arccot \frac{\cot x + \frac{\alpha}{2}}{\delta}$ & $\delta$ & $0$ & $0 < u <
       \pi$  \\[0.2cm]
  \hline\hline
\end{tabular}

\end{table}

\begin{table}[h!]

\caption{Potentials $U(u)$ and additive constants $c$.}\label{table7}
\vspace{1mm}

\centering
\begin{tabular}{lll}
  \hline\hline\\[-0.2cm]
  Type & $U(u)$ & $c$  \\[0.2cm]
  \hline\\[-0.2cm]
  SHO$_a$ & $A(A-1) \csc^2 u + 2B \cot u$ & $(\alpha - 2\delta^2) \frac{\omega^2}{4\alpha^2} +
       \frac{\beta}{\alpha} d\omega + d^2$  \\[0.2cm]
  & $A = \frac{\lambda}{\alpha}$, $B = - \frac{\omega}{2\alpha\delta} \left(d + \frac{\beta\omega}{2\alpha}
       \right)$  \\[0.2cm]
  SHO$_b$ & $A(A-1) \sec^2 u$ & $ - \frac{\omega^2 - 4\alpha d^2}{4\alpha}$  \\[0.2cm]
  & $A = \frac{\lambda}{\alpha}$   \\[0.2cm]
  RHO & $[A(A-1) + B^2] \sec^2 u - B(2A-1) \sec u \tan u$ & $- \alpha \left(L(L+1) + \frac{\omega^2}
       {4\alpha^2}\right)$  \\[0.2cm]
  & $A = \frac{1}{2} \left(L+1 + \frac{\mu}{\alpha}\right)$, $B = \frac{1}{2} \left(L+1 - \frac{\mu}{\alpha}
       \right)$  \\[0.2cm]
  C & $A(A-1) \csch^2 u - 2B \coth u$ & $\frac{\alpha}{2} [2Z + \alpha L(L+1)]$  \\[0.2cm]
  & $A = L+1$, $B = \frac{2Z}{\alpha} + L(L+1)$  \\[0.2cm]
  M & $A'(A'-1) \csch^2 u - 2B' \coth u$ & $\frac{1}{2\alpha^2} B [B + \alpha(2A+1)]$  \\[0.2cm]
  & $A' = - \frac{\lambda}{\alpha}$, $B' = \frac{1}{\alpha^2} B [B + \alpha(2A+1)]$  \\[0.2cm]
  E$_a$ & $A(A-1) \csch^2 u - 2B' \coth u$ & $\alpha [B + \delta A(A-1)]$  \\[0.2cm]
  & $B' = \frac{1}{\delta} \left(B + \frac{1}{2}\alpha A(A-1)\right)$  \\[0.2cm]
  E$_b$ & $- \frac{2Z}{u} + \frac{L(L+1)}{u^2}$ & $- 2B$  \\[0.2cm]
  & $Z = B - A(A-1)$, $L = A-1$  \\[0.2cm]
  PT & $A'(A'-1) \sec^2 u$ & $\frac{\alpha A(A-1)}{1+\alpha}$  \\[0.2cm]
  & $A' = \frac{\lambda}{1 + \alpha}$  \\[0.2cm]
  S & $[A'(A'-1) + B'^2] \sec^2 u - B'(2A'-1) \sec u \tan u$ & $- \frac{\alpha}{1 - \alpha^2}\{\alpha [A(A-1)
       + B^2]$  \\[0.2cm]
  & $A' = \frac{1}{2}(1 + \Delta_1 + \Delta_2)$, $B' = \frac{1}{2}(\Delta_2 - \Delta_1)$ & $\: \mbox{} +
       B(2A-1)\}$ \\[0.2cm]
  & $\Delta_1 = \frac{\Delta_-}{1+\alpha}$, $\Delta_2 = \frac{\Delta_+}{1-\alpha}$ \\[0.2cm]
  RM & $A(A-1) \csc^2 u + 2B' \cot u$ & $\left(\frac{\alpha^2}{2} - \beta\right) A(A-1) - \alpha B$
      \\[0.2cm]
  & $B' = \frac{1}{\delta} \left[B - \frac{1}{2} \alpha A(A-1)\right]$  \\[0.2cm]
  \hline\hline
\end{tabular}
\end{table}

\begin{table}[h!]

\caption{Bound-state wavefunctions resulting from the PCT approach.}\label{table8}
\vspace{1mm}

\centering
\begin{tabular}{lll}
  \hline\hline\\[-0.2cm]
  Type & $\psi_n^{(\alphab)}(x)$ & $t$  \\[0.2cm]
  \hline\\[-0.2cm]
  SHO$_a$ & $f^{- \frac{1}{2}(A+n+1)} \exp\left(- \frac{|B|}{A+n} \arccot \frac{\alpha x + \beta}{\delta}
       \right) R_n^{\left(\frac{2|B|}{A+n}, - A - n + 1\right)}(t)$ & $\frac{\alpha x + \beta}{\delta}$  \\[0.2cm]
  SHO$_b$ & $f^{- \frac{1}{2}(A+1)} C_n^{(A)}(t)$ & $\sqrt{\frac{\alpha}{f}} \left(x - \frac{2d}{\omega}
       \right)$  \\[0.2cm]
  RHO & $x^{A+B} f^{- A - \frac{1}{2}} P_n^{\left(A-B-\frac{1}{2}, A+B-\frac{1}{2}\right)}(t)$ & $\frac{-1
       + \alpha x^2}{1 + \alpha x^2}$  \\[0.2cm]
  C & $x^{A+n} f^{- \frac{1}{2} \left(A+n+1+\frac{B}{A+n}\right)} P_n^{\left(-A-n+\frac{B}{A+n}, -A-n
       -\frac{B}{A+n}\right)}(t)$ & $\frac{2+\alpha x}{\alpha x}$  \\[0.2cm]
  M & $f^{- \frac{1}{2} \left(A'+n+1+\frac{B'}{A'+n}\right)} \exp\left[\frac{1}{2}\left(A'+n-\frac{B'}{A'+n}
       \right) x \right]$ & $1 + 2\alpha e^{-x}$  \\[0.2cm]
  & $ \: \mbox{} \times P_n^{\left(-A'-n+\frac{B'}{A'+n}, -A'-n-\frac{B'}{A'+n}\right)}(t)$  \\[0.2cm]
  E$_a$ & $(\coth x + 1)^{\frac{1}{2}} (\coth x + 1 + \alpha)^{- \frac{1}{2} \left(A+n+1+\frac{B'}{A+n}\right)}
       $ & $\frac{2\coth x + \alpha}{2\delta}$  \\[0.2cm]
  & $ \: \mbox{} \times (\coth x - 1)^{- \frac{1}{2} \left(A+n-\frac{B'}{A+n}\right)} P_n^{\left(-A-n+\frac{B'}
       {A+n}, -A-n-\frac{B'}{A+n}\right)}(t)$  \\[0.2cm]
  E$_b$ & $(\coth x + 1)^{\frac{1}{2}} (\coth x - 1)^{-L-\frac{3}{2}} \exp\left(- \frac{Z}{(n+L+1)(\coth x - 1)}
       \right)$ & $\frac{2Z}{(n+L+1)(\coth x - 1)}$  \\[0.2cm]
  & $\: \mbox{} \times L_n^{(2L+1)}(t)$ \\[0.2cm]
  PT & $f^{- \frac{1}{2}(A'+1)} (\cos x)^{A'} C_n^{(A')}(t)$ & $\sqrt{\frac{1+\alpha}{f}} \sin x$ \\[0.2cm]
  S & $f^{- \frac{1}{2}(\Delta_1 + \Delta_2 + 2)} (1-\sin x)^{\frac{1}{2}\left(\Delta_1 + \frac{1}{2}\right)}
      (1+\sin x)^{\frac{1}{2}\left(\Delta_2 + \frac{1}{2}\right)}$ & $\frac{\sin x + \alpha}{1 + \alpha \sin x}$
       \\[0.2cm]
  & $\: \mbox{} \times  P_n^{(\Delta_1, \Delta_2)}(t)$  \\[0.2cm]
  RM & $f^{- \frac{1}{2}(A+n+1)} (\sin x)^{A+n} \exp\left(\frac{B'}{A+n} \arccot \frac{\cot x +
      \frac{\alpha}{2}}{\delta}\right)$ & $\frac{2\cot x + \alpha}{2\delta}$  \\[0.2cm]
  & $\: \mbox{} \times  R_n^{\left(-\frac{2B'}{A+n}, - A - n + 1\right)}(t)$  \\[0.2cm]
  \hline\hline
\end{tabular}
\end{table}

In Table~\ref{table8}, we remark that in addition to Laguerre, Gegenbauer, and Jacobi polynomials, there appear the less known Romanovski polynomials $R_n^{(\alpha, \beta)}(t)$ \cite{romanovski} for the shifted harmonic oscillator and the Rosen--Morse I potentials. We indeed follow here a recent approach to the wavefunctions of the latter potential for constant mass~\cite{compean}, which has stressed the interest of employing such polynomials, solutions of the second-order dif\/ferential equation
\begin{gather*}
  \left(\left(1+t^2\right) \frac{d^2}{dt^2} + (2\beta t + \alpha) \frac{d}{dt} - n (n-1+2\beta)\right) R_n^{(\alpha, \beta)}
  (t) = 0,
\end{gather*}
instead of Jacobi polynomials with complex indices and complex arguments, as is usual \cite{cooper}. For more details on the properties of Romanovski polynomials, we refer the reader to \cite{raposo}.

Finally, in Table~\ref{table9}, we provide the connections between the wavefunctions listed in Tables~\ref{table4} and~\ref{table8}. To establish those results, we have used not only equations~(\ref{eq:gegenbauer}) and~(\ref{eq:jacobi-2}), but also a~combination
\begin{gather}
  (n+1)(\alpha+n) L_{n+1}^{\alpha-2}(t) = \left((\alpha-1) t \frac{d}{dt} + \alpha(\alpha-1) - (\alpha+n) t
  \right) L_n^{(\alpha)}(t)  \label{eq:laguerre}
\end{gather}
of the recursion and dif\/ferential relations of Laguerre polynomials (see equations (8.971.5) and (8.971.3) of \cite{gradshteyn}), the backward shift operator relation for Jacobi polynomials (see equation~(1.8.7) of \cite{koekoek}),
\begin{gather}
  2(n+1) P_{n+1}^{(\alpha, \beta)}(t) = \left(- \left(1-t^2\right) \frac{d}{dt} + \alpha - \beta + (\alpha+\beta+2) t\right)
  P_n^{(\alpha+1, \beta+1)}(t),  \label{eq:jacobi-1}
\end{gather}
and the dif\/ferential relation of Romanovski polynomials (see equation~(41) of \cite{raposo}),
\begin{gather}
  \frac{n+2\beta-1}{2(n+\beta)} R_{n+1}^{(\alpha, \beta)}(t) = \left[\left(1+t^2\right) \frac{d}{dt} + (n+2\beta-1)
  \left(t + \frac{\alpha}{2(n+\beta)}\right)\right] R_n^{(\alpha, \beta)}(t).  \label{eq:romanovski}
\end{gather}

\begin{table}[h!]

\caption{Changes of variable and of function relating the wavefunctions arising from the DSI and PCT approaches and equations used in the comparison.}\label{table9}
\vspace{1mm}

\centering
\begin{tabular}{llll}
  \hline\hline\\[-0.2cm]
  Type & $t$ & $P_n(\lambdab; y)$ & Equation  \\[0.2cm]
  \hline\\[-0.2cm]
  SHO$_a$ & $\frac{\alpha y + \beta}{\delta}$ & $\gamma_n^{(A)} R_n^{\left(\frac{2|B|}{A+n}, -A-n+1\right)}
       (t)$ & (\ref{eq:romanovski})  \\[0.2cm]
  SHO$_b$ & $\sqrt{\frac{\alpha}{f}} \left(y - \frac{2d}{\omega}\right)$ & $\gamma_n^{(A)} f^{\frac{n}{2}}
       C_n^{(A)}(t)$ & (\ref{eq:gegenbauer})  \\[0.2cm]
  RHO & $\frac{-1+\alpha y}{1+\alpha y}$ & $\gamma_n f^n P_n^{\left(A-B-\frac{1}{2}, A+B-\frac{1}{2}
       \right)}(t)$ & (\ref{eq:jacobi-1})  \\[0.2cm]
  C & $\frac{2y+\alpha}{\alpha}$ & $\gamma_n^{(A)} P_n^{\left(-A-n+\frac{B}{A+n}, -A-n-\frac{B}{A+n}
       \right)}(t)$ & (\ref{eq:jacobi-2})  \\[0.2cm]
  M & $1+2\alpha y$ & $\gamma_n^{(A')} P_n^{\left(-A'-n+\frac{B'}{A'+n}, -A'-n-\frac{B'}{A'+n}\right)}(t)$ &
       (\ref{eq:jacobi-2})  \\[0.2cm]
  E$_a$ & $\frac{2y+\alpha}{2\delta}$ & $\gamma_n^{(A)} P_n^{\left(-A-n+\frac{B'}{A+n}, -A-n-
       \frac{B'}{A+n}\right)}(t)$ & (\ref{eq:jacobi-2})  \\[0.2cm]
  E$_b$ & $- \frac{2Z}{(n+L+1)(1-y)}$ & $\gamma_n^{(L)} t^{-n} L_n^{(2L+1)}(t)$ & (\ref{eq:laguerre})
       \\[0.2cm]
  PT & $y \sqrt{\frac{1+\alpha}{1+(1+\alpha)y^2}}$ & $\gamma_n^{(A')} (1-t^2)^{-\frac{n}{2}}
       C_n^{(A')}(t)$ & (\ref{eq:gegenbauer})  \\[0.2cm]
  S & $\frac{y+\alpha}{1+\alpha y}$ & $\gamma_n f^n P_n^{(\Delta_1, \Delta_2)}(t)$ & (\ref{eq:jacobi-1})
       \\[0.2cm]
  RM & $\frac{2y+\alpha}{2\delta}$ & $\gamma_n^{(A)} R_n^{\left(\frac{-2B'}{A+n}, -A-n+1\right)}(t)$ &
       (\ref{eq:romanovski})  \\[0.2cm]
  \hline\hline
\end{tabular}
\end{table}

\section{Conclusion}\label{section4}

In the present paper, we have completed the study of a class of exactly solvable PDM Schr\"odinger equations, undertaken in \cite{bagchi05a} and pursued in~\cite{cq08, cq07}, by constructing their bound-state wavefunctions in general form. This has been made possible by combining the previously used DSI method with PCT connecting the PDM Schr\"odinger equations with constant-mass ones with similar spectra.

It should be stressed that it is only the association of both approaches that has allowed us to build and to fully solve the former equations. As shown in Table~\ref{table6}, the complexity of the changes of variable involved in the PCT would indeed make it rather unlikely to guess them in order to directly construct the PDM equations from the constant-mass ones. The previous determination of the PDM and of the corresponding spectra by DSI techniques has therefore played an essential role.

\newpage

\pdfbookmark[1]{References}{ref}
\LastPageEnding

\end{document}